# Statistics of the first passage area functional for an Ornstein-Uhlenbeck process

Michael J Kearney [1] and Richard J Martin [2]

[1] Senate House, University of Surrey, Guildford, Surrey GU2 7XH, UK

[2] Department of Mathematics, Imperial College London,
South Kensington, London, SW7 2AZ, UK

*Abstract*

We consider the area functional defined by the integral of an Ornstein-Uhlenbeck process which starts from a given value and ends at the time it first reaches zero (its equilibrium level). Exact results are presented for the mean, variance, skewness and kurtosis of the underlying area probability distribution, together with the covariance and correlation between the area and the first passage time. Amongst other things, the analysis demonstrates that the area distribution is asymptotically normal in the weak noise limit, which stands in contrast to the first passage time distribution. Various applications are indicated.

# 1. Introduction

First passage problems play a special role in the theory of stochastic processes [1]. The study of first passage functionals of Brownian motion [2] is an important sub-discipline and one which finds many applications. This includes the analysis of lattice polygons and queueing systems [3,4], snowmelt dynamics [5], DNA breathing [6], polymer translocation [7] and barrierless reaction kinetics [8]. Conditioning on a fixed first passage time provides insight into interface fluctuations [9], laser cooling of atoms [10] and colloidal suspensions [11]; the associated techniques being developed explicitly in [12, 13], along with studies of functional correlations in [14, 15]. A recent overview of key results, which also provides some important new insights into asymptotic behaviour, may be found in [16].

In this paper, we are interested in the first passage area functional associated with the relaxation of an Ornstein-Uhlenbeck process to zero (its equilibrium level, also known as the mean reversion level). The area functional $A$ is defined by the time integral of the process starting from a given initial value up to the first passage time $T$ to reach zero. Both $A$ and $T$ are well-defined random variables. The former arises, either directly or through natural extension, in the modelling of two-dimensional diffusion [17], random acceleration [18-20], inelastic collapse [21], discrete-time persistence [22] and neuron dynamics [23, 24]. An understanding of the statistics of this area variable is therefore relevant to a variety of different problems.

A specific example which is easy to visualise is that of a queueing system which features balking. This term refers to when 'customers' arriving at a queue decide not to join because the queue is too long. The context can be quite general; for example, an everyday example is the road traffic jam, where balking refers to when forewarned drivers find an alternative route instead of joining the jam. At large scale the Ornstein-Uhlenbeck process is a good approximation to the evolution of the queue length of a critically balanced queue when the rate of balking is proportional to the queue length [25-27]. One then identifies $T$ as the duration (or lifetime) of the queue and $A$ as the total waiting time of all the customers involved in the queue during its lifetime, an important operational measure of overall inefficiency or 'cost'.

The first passage time probability density is known exactly for the specific problem in hand [28-30]. It is unlikely, however, that the area probability density has a simple, closed-form solution. To make progress, here we derive exact expressions for the mean, variance, skewness and kurtosis of $A$. The behaviour in the weak noise limit is of special interest and we shall establish that the law of $A$ is asymptotically normal (Gaussian) in this limit. This contrasts with the law of $T$ which is not asymptotically normal. An explanation as to why this convergence to Gaussian behaviour occurs, in terms of the generic properties of integrated Gaussian processes, is given in the Discussion section at the end.

The reasons why Gaussian or non-Gaussian behaviour arises in a given stochastic process is of wider interest; see e.g. [31]. A topical example, which also involves the study of integrated Ornstein-Uhlenbeck processes and consideration of functionals of Brownian motion, relates to so-called diffusing diffusivity models. These have been invoked to explain the behaviour of diffusion-type processes which have a linear time dependence of the mean-square displacement, but accompanied by a non-Gaussian displacement distribution, possibly crossing over to Gaussian on long timescales. An early approach modelled the diffusivity as a biased random walk [32]. Subsequent refinements have included treating the diffusivity as the square of an Ornstein-Uhlenbeck process, and formally introducing the subordination concept, which associates a random variable with the number of steps of some underlying stochastic process [33, 34]. Describing the fluctuating diffusivity through subordination highlights the importance of the probability density of the integrated diffusivity (a functional), which lends itself to even more generalised treatments, including making full use of spectral concepts [35, 36]. The link between this and the current work is only indirect, but the broader contextual setting is interesting, and the results and techniques developed here may find wider use.

The paper is structured as follows. In Section 2, the basic theory is developed, culminating in an explicit expression for the double Laplace transform of the joint probability density of $A$ and $T$. This extends the known results for the Laplace transforms of the probability densities of $A$ and $T$ when considered in isolation. In Section 3, for the purposes of comparison, we summarise the key results relating to the first passage time $T$. As mentioned above, the probability density of $T$ is known, but the explicit results given for the asymptotic behaviour of the moments and cumulants are essentially new. In Section 4, the main results relating to $A$ are presented, principally concerning the moments and cumulants, together with analysis of the correlation between $A$ and $T$. Except for the first moment of $A$, all these results are new to the best of our knowledge. Finally, in Section 5, the findings are discussed, and conclusions are drawn.

## 2. General theory

The process we wish to study can be modelled by the Langevin equation

$$\frac{dy(t)}{dt} = -\gamma_d y(t) + \sigma \xi(t); \qquad y(t=0) = y_0 > 0. \qquad (1a)$$

Here, $\xi(t)$ is a Gaussian white noise source such that $\xi(t)dt \equiv dW(t)$, with $W(t)$ denoting the Wiener process. The first passage time to reach zero is defined by $T \equiv \min\{t : y(t) = 0\}$, whilst the area functional is defined by $A = \int_0^T y(t)\,dt$. It is convenient to work in dimensionless units via the transformation

$$\tau = \gamma_d t; \qquad x = \sqrt{\frac{2\gamma_d}{\sigma^2}} y; \qquad \mathcal{T} = \gamma_d T; \qquad \mathcal{A} = \sqrt{\frac{2\gamma_d^3}{\sigma^2}} A$$

which gives

$$\frac{dx(\tau)}{d\tau} = -x(\tau) + \sqrt{2}\xi(\tau); \qquad x(\tau=0) = \alpha \qquad (1b)$$

where $\alpha \equiv \sqrt{2\gamma_d/\sigma^2}\, y_0$, with $\mathcal{T} \equiv \min\{\tau : x(\tau) = 0\}$ and $\mathcal{A} = \int_0^{\mathcal{T}} x(\tau)\,d\tau$. All statistical quantities relating to $\mathcal{T}$ and $\mathcal{A}$ are functions of $\alpha$ only.

Rather than study the variable $\mathcal{A}$ in isolation, there is merit in considering the joint probability density $P(\mathcal{A},\mathcal{T}|\alpha)$, in terms of which one can express the respective individual densities by integrating out the redundant variable. We start by considering the double Laplace transform of $P(\mathcal{A}, \mathcal{T} | \alpha)$:

$$\tilde{P}(s, p, \alpha) = \int_0^\infty \int_0^\infty P(\mathcal{A}, \mathcal{T} | \alpha) e^{-s\mathcal{A}} e^{-p\mathcal{T}} \, \mathrm{d}\mathcal{A} \, \mathrm{d}\mathcal{T} \equiv \mathrm{E}[e^{-s\mathcal{A}} e^{-p\mathcal{T}}]. \quad (2)$$

This quantity satisfies a backward Fokker-Planck equation

$$\left[ \frac{\mathrm{d}^2}{\mathrm{d}\alpha^2} - \alpha \frac{\mathrm{d}}{\mathrm{d}\alpha} - (s\alpha + p) \right] \tilde{P}(s, p, \alpha) = 0 \quad (3)$$

subject to the boundary conditions $\tilde{P}(s, p, \alpha = 0) = 1$ and $\tilde{P}(s, p, \alpha \to \infty) = 0$. To show this, one can adapt the 'path-integral' procedure outlined in [2, 3], starting by rewriting (2) as follows:

$$\tilde{P}(s, p, \alpha) = \left\langle \exp\left[ -\int_0^{\mathcal{T}} (sx(\tau) + p) \, \mathrm{d}\tau \right] \right\rangle$$

where $\langle \ldots \rangle$ implies the average over all eligible realisations of $x(\tau)$. A path of the process over the interval $[0, \mathcal{T}]$ may be split into two parts: a left interval $[0, \Delta\tau]$ where the process proceeds from $x(0) \equiv \alpha$ to $\alpha + \Delta x$ in a small time $\Delta\tau$, and a right interval $[\Delta\tau, \mathcal{T}]$ where the process starts from $\alpha + \Delta x$ at time $\Delta\tau$ and reaches zero at time $\mathcal{T}$. Noting that $\int_0^{\Delta\tau} (sx(\tau) + p) \mathrm{d}\tau \approx (s\alpha + p)\Delta\tau$ gives

$$\tilde{P}(s, p, \alpha) = \left\langle e^{-(s\alpha + p)\Delta\tau} \tilde{P}(s, p, \alpha + \Delta x) \right\rangle_{\Delta x} + \ldots$$

where the average is over all possible realisations of $\Delta x$. Expanding the right-hand side to $O(\Delta x^2)$ and using the fact that $\langle \Delta x \rangle \approx -\alpha \Delta \tau$ and $\langle \Delta x^2 \rangle \approx 2\Delta \tau$, which follows from (1b), one obtains (3). Inspection of (2) shows that $\tilde{P}(s=0,p,\alpha)$ is the Laplace transform of the first passage time density $P(\mathcal{T}|\alpha)$; whilst $\tilde{P}(s,p=0,\alpha)$ is the Laplace transform of the area density $P(\mathcal{A}|\alpha)$.

The formal solution of (3) is

$$\tilde{P}(s,p,\alpha) = e^{\frac{1}{4}\alpha^2} \frac{D_{-p+s^2}(2s+\alpha)}{D_{-p+s^2}(2s)}. \tag{4}$$

Here, $D_\nu(z)$ is a parabolic cylinder (Weber) function which satisfies [37]

$$\frac{d^2 D_\nu(z)}{dz^2} + \left[\nu + \frac{1}{2} - \frac{z^2}{4}\right] D_\nu(z) = 0; \qquad D_\nu(0) = \frac{2^{\frac{1}{2}\nu}\sqrt{\pi}}{\Gamma\left(\frac{1-\nu}{2}\right)}.$$

It is possible to invert (4) when $s=0$ to derive $P(\mathcal{T}|\alpha)$, which is given in Section 3. Unfortunately, it seems quite impractical to invert (4) when $p=0$ to derive $P(\mathcal{A}|\alpha)$, at least in any simple form [21]. One can, however, make progress in evaluating various statistical quantities without carrying out the inversion. For example, the generating functions for the basic moments are given by

$$M_{\mathcal{T}}(t,\alpha) \equiv \mathrm{E}[e^{t\mathcal{T}}] = \tilde{P}(s=0,-t,\alpha) = \sum_{n=0}^{\infty} \mathcal{T}_n(\alpha) \frac{t^n}{n!}$$

(5)

$$M_{\mathcal{A}}(t,\alpha) \equiv \mathrm{E}[e^{t\mathcal{A}}] = \tilde{P}(-t, p=0, \alpha) = \sum_{n=0}^{\infty} \mathcal{A}_n(\alpha) \frac{t^n}{n!}$$

where the connection to the Laplace transform follows from the structure of (2). In pursuing this approach, the following representation for $D_\nu(z)$ which holds for $\mathrm{Re}(\nu) < 0$ is particularly useful [37]:

$$D_\nu(z) = \frac{e^{-\frac{1}{4}z^2}}{\Gamma(-\nu)} \int_0^\infty u^{-\nu-1} e^{-\frac{1}{2}u^2 - zu} \, du.$$

Using this, one can simplify (4) for $\mathrm{Re}(p-s^2) > 0$:

$$\tilde{P}(s,p,\alpha) = e^{-\alpha s} \frac{\int_0^\infty u^{p-s^2-1} e^{-\frac{1}{2}u^2 - 2su - \alpha u} \, du}{\int_0^\infty u^{p-s^2-1} e^{-\frac{1}{2}u^2 - 2su} \, du}$$

$$= e^{-\alpha s} \left[ 1 - \frac{\int_0^\infty u^{p-s^2-1} e^{-\frac{1}{2}u^2 - 2su} (1 - e^{-\alpha u}) \, du}{\int_0^\infty u^{p-s^2-1} e^{-\frac{1}{2}u^2 - 2su} \, du} \right]$$

(6)

$$= e^{-\alpha s} \left[ 1 - (p-s^2) F(s,p,\alpha) \right]$$

where, after integrating the lower integral by parts, one has

$$F(s,p,\alpha) = \frac{\int_0^\infty u^{p-s^2-1} e^{-\frac{1}{2}u^2 - 2su} (1 - e^{-\alpha u}) \, du}{\int_0^\infty u^{p-s^2} (u+2s) e^{-\frac{1}{2}u^2 - 2su} \, du}.$$

(7)

Written in this way, the singular behaviour as $s, p \to 0$ has been regularised, making subsequent calculations somewhat easier. Much of what is to come is based on manipulating (6, 7) to derive formal power series whose coefficients are the statistical quantities of interest. In this context, setting $p = 0$ in (6, 7) is justified.

**3. The first passage time statistics**

We begin by highlighting certain key details of the statistics of the first passage time variable $\mathcal{T}$. The inversion of (4) when $s = 0$ is known exactly, although the derivation is usually by other means [28-30]:

$$P(\mathcal{T}|\alpha) = \sqrt{\frac{2}{\pi}} \frac{\alpha e^{-\mathcal{T}}}{(1-e^{-2\mathcal{T}})^{3/2}} \exp\left\{-\frac{\alpha^2 e^{-2\mathcal{T}}}{2(1-e^{-2\mathcal{T}})}\right\}. \tag{8}$$

An important feature of this result is the essential singularity as $\mathcal{T} \to 0$. One can write the moment generating function based on (5, 6) as

$$M_{\mathcal{T}}(t,\alpha) = 1 + \frac{t2^{\frac{1}{2}t}}{\Gamma(1-\tfrac{1}{2}t)} \int_0^\infty u^{-t} e^{-\frac{1}{2}u^2} \left[\frac{1-e^{-\alpha u}}{u}\right] du \tag{9}$$

From this, one can expand to derive the following integral expressions for the first two moments:

$$\mathcal{T}_1(\alpha) = \int_0^\infty e^{-\frac{1}{2}u^2} \left[ \frac{1 - e^{-\alpha u}}{u} \right] du \qquad (10)$$

$$\mathcal{T}_2(\alpha) = -2\int_0^\infty e^{-\frac{1}{2}u^2} \left[ \frac{1 - e^{-\alpha u}}{u} \right] \log u \, du + (\log 2 - \gamma)\mathcal{T}_1(\alpha) \qquad (11)$$

where $\gamma = 0.577...$ is Euler's constant. As mentioned in the Introduction, an important point of focus is the behaviour in the weak noise limit, which corresponds to $\alpha \to \infty$. Based on (A4, A5) from the Appendix one has

$$\mathcal{T}_1(\alpha \to \infty) = \log \alpha + \frac{\log 2 + \gamma}{2} + o(1) \qquad (12)$$

$$\mathcal{T}_2(\alpha \to \infty) = \left( \log \alpha + \frac{\log 2 + \gamma}{2} \right)^2 + \frac{\pi^2}{8} + o(1). \qquad (13)$$

Importantly, (12, 13) imply that the variance of the first passage time is bounded above by $\pi^2/8 = 1.233...$. Given $\mathcal{T}_n(\alpha) \equiv \int_0^\infty \mathcal{T}^n P(\mathcal{T}|\alpha) \, d\mathcal{T}$, these asymptotes may also be derived directly using (8).

The structure of (12, 13) foreshadows the limiting form of the higher order moments, which may be obtained as follows. As $\alpha \to \infty$, one has from (9) in conjunction with (A6) from the Appendix:

$$M_\mathcal{T}(t, \alpha \to \infty) = e^{t(\log \alpha + \frac{1}{2}\log 2 + \frac{1}{2}\gamma)} \times e^{-\frac{1}{2}t\gamma} \frac{\Gamma(1-t)}{\Gamma(1-\frac{1}{2}t)} + o(1)$$

whereupon differentiating repeatedly using the product rule gives

$$\mathcal{T}_n(\alpha \to \infty) = \sum_{k=0}^{n} \binom{n}{k} \left( \log \alpha + \frac{\log 2 + \gamma}{2} \right)^{n-k} C_k^{\mathcal{T}}(\infty) + o(1) \qquad (14)$$

where the constants $C_n^{\mathcal{T}}(\infty)$ are defined by

$$C_n^{\mathcal{T}}(\infty) \equiv \frac{\partial^n}{\partial t^n} \left[ e^{-\frac{1}{2}\gamma t} \frac{\Gamma(1-t)}{\Gamma(1-\frac{1}{2}t)} \right]_{t=0}. \qquad (15)$$

These may be evaluated with the help of the identity [37]

$$\log \Gamma(1-z) = \gamma z + \sum_{n=2}^{\infty} \frac{\zeta(n)}{n} z^n; \qquad |z| < 1$$

where $\zeta(n)$ is the Riemann zeta function. The first few values are given by

$$C_2^{\mathcal{T}}(\infty) = \frac{\pi^2}{8}; \qquad C_3^{\mathcal{T}}(\infty) = \frac{7\zeta(3)}{4}; \qquad C_4^{\mathcal{T}}(\infty) = \frac{7\pi^4}{64}; \qquad \ldots \qquad (16)$$

A little thought convinces one that these constants $C_n^{\mathcal{T}}(\infty)$ are none other than the limiting central moments, e.g., by recognising that one can write

$$\mathcal{T}_n(\alpha) \equiv \mathrm{E}\left[ \left( \mathcal{T}_1(\alpha) + (\mathcal{T} - \mathcal{T}_1(\alpha)) \right)^n \right] = \sum_{k=0}^{n} \binom{n}{k} \mathcal{T}_1(\alpha)^{n-k} C_k^{\mathcal{T}}(\alpha).$$

One can further show that there is a concise encapsulation of these results in terms of the limiting cumulants $K_n^{\mathcal{T}}(\infty)$, which are given for all $n \geq 2$ by

$$K_n^{\mathcal{T}}(\infty) = (n-1)!\zeta(n)(1-2^{-n}).$$

The limiting density is non-Gaussian; it is, in fact, an exponential gamma distribution, a relative of the Gumbel distribution, as may be shown directly from (8). The limiting form of the skewness $\gamma_{\mathcal{T}}(\alpha)$ and kurtosis $\kappa_{\mathcal{T}}(\alpha)$ as $\alpha \to \infty$ are

$$\gamma_{\mathcal{T}}(\infty) = \frac{28\sqrt{2}\zeta(3)}{\pi^3}; \qquad \kappa_{\mathcal{T}}(\infty) = 7.$$

As a point of note, a Gaussian variable has skewness 0 and kurtosis 3.

**4. The area statistics**

We now turn to the main topic of the paper, that of deriving the statistics of the area variable $\mathcal{A}$. Noting from (5) that

$$\mathcal{A}_n(\alpha) = \frac{\partial^n}{\partial t^n} M_{\mathcal{A}}(t,\alpha)\bigg|_{t=0} = (-1)^n \frac{\partial^n}{\partial s^n} \tilde{P}(s,p=0,\alpha)\bigg|_{s=0}$$

one deduces from (3) that the area moments satisfy the recursive differential equation

$$\left[\frac{d^2}{d\alpha^2} - \alpha\frac{d}{d\alpha}\right]\mathcal{A}_n(\alpha) = -n\alpha\mathcal{A}_{n-1}(\alpha); \qquad \mathcal{A}_0(\alpha) \equiv 1 \qquad (17)$$

subject to $\mathcal{A}_n(\alpha = 0) = 0$. The solution of (17) in integral form is given by the 'Siegert-type' formula [38]

$$\mathcal{A}_n(\alpha) = n\int_0^\alpha e^{\frac{1}{2}z^2} \int_z^\infty z' e^{-\frac{1}{2}z'^2} \mathcal{A}_{n-1}(z')\,dz'\,dz.$$

It follows at once that the first area moment is given by

$$\mathcal{A}_1(\alpha) = \alpha. \qquad (18)$$

This surprisingly simple result was first identified in [17]. It is interesting to note that this is what one would obtain for the corresponding deterministic problem, i.e. $x(\tau) = \alpha e^{-\tau}$, for which the first passage time $\mathcal{T} = \infty$.

Given $\mathcal{A}_1(\alpha) = \alpha$, it follows rather neatly that the generating function for the central moments $C_n^{\mathcal{A}}(\alpha)$ can be written based on (6) as

$$C_{\mathcal{A}}(t,\alpha) \equiv \mathrm{E}\left[e^{t(\mathcal{A}-\alpha)}\right] = e^{-\alpha t}\tilde{P}(-t, p=0, \alpha) = 1 + t^2 F(-t, p=0, \alpha)$$

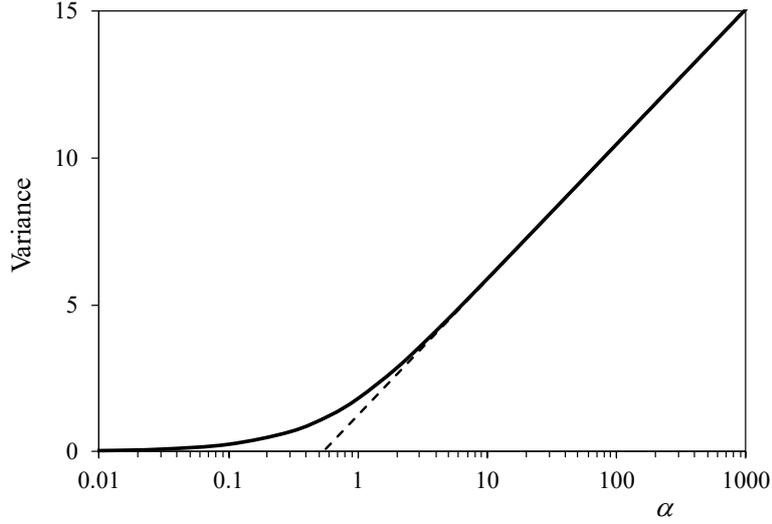

Figure 1: The variance of $\mathcal{A}$ as a function of $\alpha$ (solid line). The $\alpha \to \infty$ asymptote is indicated (dashed line).

from which one has for $n \geq 2$:

$$C_n^{\mathcal{A}}(\alpha) = n(n-1)\frac{\partial^{n-2}}{\partial t^{n-2}} F(-t, p=0, \alpha)\bigg|_{t=0}. \qquad (19)$$

For the second central moment $C_2^{\mathcal{A}}(\alpha)$, which is simply the variance, one can quickly derive the following expression based on (7, 19):

$$C_2^{\mathcal{A}}(\alpha) = 2\int_0^\infty e^{-\frac{1}{2}u^2}\left[\frac{1-e^{-\alpha u}}{u}\right]du = 2\mathcal{T}_1(\alpha). \qquad (20)$$

The last step draws on (10). This is plotted in figure 1 and the behaviour as $\alpha \to \infty$ follows from (12):

$$C_2^{\mathcal{A}}(\alpha \to \infty) = 2\log\alpha + \log 2 + \gamma + o(1).$$

In conjunction with (18), the second area moment itself is given by

$$\mathcal{A}_2(\alpha) = \alpha^2 + 2\int_0^\infty e^{-\frac{1}{2}u^2}\left[\frac{1-e^{-\alpha u}}{u}\right]du = \alpha^2 + 2\mathcal{T}_1(\alpha). \tag{21}$$

One can readily show that $\mathcal{A}_2(\alpha)$ satisfies (17).

To characterize the distribution more fully, it is useful to study the third and fourth central moments. For the former we have from (7, 19):

$$C_3^{\mathcal{A}}(\alpha) = 12\int_0^\infty e^{-\frac{1}{2}u^2}(1-e^{-\alpha u})\,du = 6\sqrt{2\pi}\left[1 - e^{\frac{1}{2}\alpha^2}\mathrm{erfc}\left(\frac{\alpha}{\sqrt{2}}\right)\right]. \tag{22}$$

The asymptotic behaviour as $\alpha \to \infty$ is easily extracted:

$$C_3^{\mathcal{A}}(\alpha \to \infty) = 6\sqrt{2\pi} + o(1).$$

For the fourth central moment the analysis is slightly more demanding. One has based on (7, 19):

$$C_4^{\mathcal{A}}(\alpha) = -24\int_0^\infty e^{-\frac{1}{2}u^2}\left[\frac{1-e^{-\alpha u}}{u}\right]\log u\, du + 48\int_0^\infty ue^{-\frac{1}{2}u^2}(1-e^{-\alpha u})\,du$$

$$+12(\log 2 - \gamma)\int_0^\infty e^{-\frac{1}{2}u^2}\left[\frac{1-e^{-\alpha u}}{u}\right]du$$

which can be written using (11) as

$$C_4^{\mathcal{A}}(\alpha) = 12\mathcal{T}_2(\alpha) + 24\sqrt{2\pi}\,\alpha e^{\frac{1}{2}\alpha^2}\operatorname{erfc}\left(\frac{\alpha}{\sqrt{2}}\right). \tag{23}$$

Based on (13) the $\alpha \to \infty$ behaviour is then given by

$$C_4^{\mathcal{A}}(\alpha \to \infty) = 12\left(\log\alpha + \frac{\log 2 + \gamma}{2}\right)^2 + \frac{3\pi^2}{2} + 48 + o(1).$$

From (20, 22, 23) one can derive the skewness $\gamma_{\mathcal{A}}(\alpha)$ and kurtosis $\kappa_{\mathcal{A}}(\alpha)$ of the underlying area probability distribution. The results are displayed in figure 2. Concerning the $\alpha \to \infty$ limit one has

$$\gamma_{\mathcal{A}}(\alpha \to \infty) = \frac{6\sqrt{2\pi}}{(2\log\alpha + \log 2 + \gamma)^{3/2}}[1 + o(1)]$$

$$\kappa_{\mathcal{A}}(\alpha \to \infty) = 3 + \frac{3\pi^2 + 96}{2(2\log\alpha + \log 2 + \gamma)^2}[1 + o(1)].$$

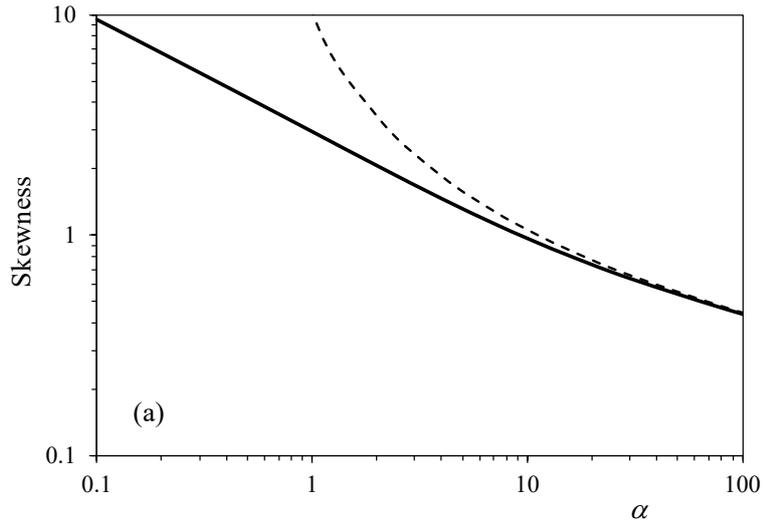

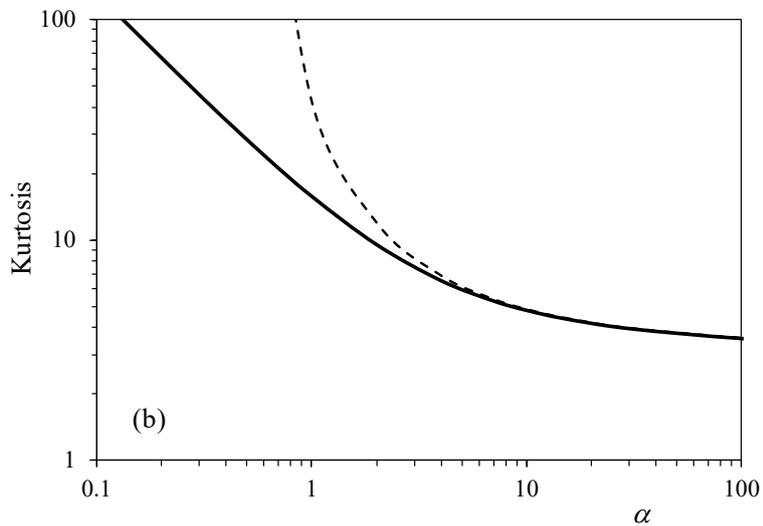

Figure 2: The skewness (a) and kurtosis (b) of $\mathcal{A}$ as a function of $\alpha$ (solid lines). The $\alpha \to \infty$ asymptotes are indicated (dashed lines).

This suggests that the area distribution is asymptotically normal as $\alpha \to \infty$, although convergence is slow. A more precise statement comes from considering the limiting form of the area cumulants $K_n^{\mathcal{A}}(\alpha)$, whose generating function based on (6) may be written as:

$$K_{\mathcal{A}}(t,\alpha) = \log \tilde{P}(-t, p=0, \alpha) = \alpha t + \log\left[1 + t^2 F(-t, p=0, \alpha)\right].$$

One can then show using (7) and (A6) from the Appendix that

$$K_{\mathcal{A}}(t, \alpha \to \infty) = \alpha t + t^2 \log \alpha + \log\left[\frac{\Gamma(1-t^2)}{f(t)}\right] + o(1)$$

$$= \alpha t + t^2(\log \alpha + \gamma) + \sum_{k=2}^{\infty} \frac{\zeta(k)}{k} t^{2k} - \log f(t) + o(1)$$

(24)

where

$$f(t) = \int_0^\infty u^{-t^2}(u - 2t) e^{-\frac{1}{2}u^2 + 2tu}\, du.$$

The function $f(t)$ does not have a simple expansion as a power series, but the first few terms based on a direct calculation are

$$f(t) = 1 - \left(\frac{\log 2 - \gamma}{2}\right)t^2 - \sqrt{2\pi}\, t^3 + \left[\frac{1}{2}\left(\frac{\log 2 - \gamma}{2}\right)^2 + \frac{\pi^2}{48} - 2\right]t^4 + \ldots$$

whereupon

$$K_{\mathcal{A}}(t, \alpha \to \infty) = \alpha t + \frac{t^2}{2!}(2\log \alpha + \log 2 + \gamma)$$

$$+ \frac{t^3}{3!} 6\sqrt{2\pi} + \frac{t^4}{4!}\left(\frac{3\pi^2}{2} + 48\right) + \ldots.$$

(25)

The first two terms of (25) are trivial, since $K_1^{\mathcal{A}}(\alpha) = \mathcal{A}_1(\alpha)$ and $K_2^{\mathcal{A}}(\alpha) = C_2^{\mathcal{A}}(\alpha)$.

The important observation, as determined from (24), is that all cumulants $K_n^{\mathcal{A}}(\alpha)$ for $n \geq 3$ tend to constant values as $\alpha \to \infty$, e.g.,

$$K_3^{\mathcal{A}}(\infty) = 6\sqrt{2\pi}; \qquad K_4^{\mathcal{A}}(\infty) = \frac{3\pi^2}{2} + 48; \qquad \ldots$$

Based on the following well-known relationship between cumulants and central moments:

$$C_n^{\mathcal{A}}(\alpha) = \sum_{k=2}^{n} \binom{n-1}{k-1} K_k^{\mathcal{A}}(\alpha) C_{n-k}^{\mathcal{A}}(\alpha)$$

this means that for the limiting central moments one has

$$C_{2n}^{\mathcal{A}}(\alpha \to \infty) = \frac{(2n)!}{n!}\left(\log\alpha + \frac{\log 2 + \gamma}{2}\right)^n [1 + o(1)]$$

$$C_{2n+1}^{\mathcal{A}}(\alpha \to \infty) = \sqrt{2\pi}\,\frac{(2n+1)!}{(n-1)!}\left(\log\alpha + \frac{\log 2 + \gamma}{2}\right)^{n-1} [1 + o(1)].$$

(26)

These results confirm that the area distribution is asymptotically normal, in the sense that there is a suitably rescaled area variable $\hat{\mathcal{A}}$ such that

$$\hat{\mathcal{A}} \equiv \frac{\mathcal{A} - \alpha}{\sqrt{2\log\alpha + \log 2 + \gamma}} \sim N(0,1); \qquad \alpha \to \infty \qquad (27)$$

which is based on the reduction of (25, 26) to the Gaussian forms

$$K_{\hat{\mathcal{A}}}(t,\alpha) \to \tfrac{1}{2}t^2; \qquad C_{2n}^{\hat{\mathcal{A}}}(\alpha) \to (2n-1)!!; \qquad C_{2n+1}^{\hat{\mathcal{A}}}(\alpha) \to 0.$$

This is different from the first passage time distribution which is not asymptotically normal, as demonstrated in the previous section.

Having established what we set out to achieve, we conclude by considering the matter of correlations. One naturally expects a positive correlation between $\mathcal{A}$ and $\mathcal{T}$ on physical grounds; typically, the larger the first passage time is, the larger the corresponding area will be. One can study this based on (2) and using the Laplace transform (6) to derive the covariance $\mathrm{Cov}(\mathcal{A}, \mathcal{T}) \equiv \mathrm{E}[\mathcal{A}\mathcal{T}] - \mathrm{E}[\mathcal{A}]\mathrm{E}[\mathcal{T}]$:

$$\mathrm{Cov}(\mathcal{A}, \mathcal{T}) = (-1)^2 \frac{\partial^2}{\partial s \partial p} \tilde{P}(s, p, \alpha)\bigg|_{s=p=0} - \alpha \mathcal{T}_1(\alpha)$$

$$= 2\int_0^\infty e^{-\frac{1}{2}u^2}(1 - e^{-\alpha u})\,\mathrm{d}u = \sqrt{2\pi}\left[1 - e^{\frac{1}{2}\alpha^2}\mathrm{erfc}\left(\frac{\alpha}{\sqrt{2}}\right)\right].$$

(28)

In turn, one can define the correlation coefficient

$$R(\mathcal{A}, \mathcal{T}) \equiv \frac{\mathrm{Cov}(\mathcal{A}, \mathcal{T})}{\sqrt{\mathrm{Var}(\mathcal{A})\mathrm{Var}(\mathcal{T})}}.$$

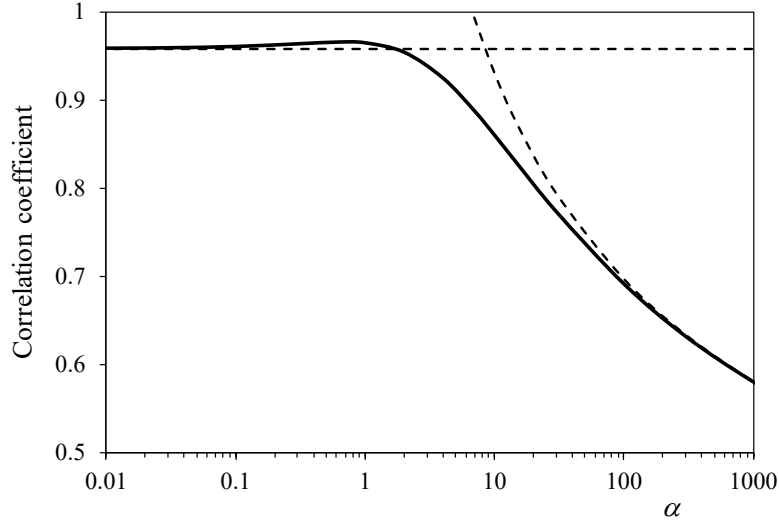

Figure 3: The correlation coefficient as a function of $\alpha$ (solid line). The $\alpha \to \infty$ and $\alpha \to 0$ asymptotes are indicated (dashed lines).

All the elements making up this expression have been derived earlier; the variance of the first passage time distribution can be computed from (10, 11). The result is presented as a function of $\alpha$ in figure 3. In terms of limiting behaviour:

$$R(\mathcal{A}, \mathcal{T}) = \frac{4}{\sqrt{\pi(2\log\alpha + \log 2 + \gamma)}}[1 + o(1)]; \qquad \alpha \to \infty$$

$$R(\mathcal{A}, \mathcal{T}) = \frac{2}{\sqrt{2\pi\log 2}} = 0.958...; \qquad \alpha \to 0.$$

The maximum is found to be $R_{max}(\mathcal{A}, \mathcal{T}) = 0.965...$ when $\alpha = 0.698...$. These results form an interesting counterpart to the corresponding results for drifted Brownian motion [14], with the observation that the strong noise limit $\alpha \to 0$ is *not* the same in both cases. The decay to zero as $\alpha \to \infty$ is a signal that, unlike the first passage time variable, the area variable is not so tightly clustered around its mean value.

**5. Discussion**

We have derived exact results for the key statistical measures relating to the distribution of the area integral of the Ornstein-Uhlenbeck process up to the first passage time to reach its equilibrium level. In addition, the correlation function between the area and the first passage time has been computed. The results should find application as they stand. One could extend the analysis relatively easily to discuss the area statistics when the first passage time relates to a level which differs from the equilibrium level. This has been done in some detail for the statistics of the first passage time itself; for a recent overview see [39-41]. More generally, e.g. with reference to the diffusing diffusivity models mentioned in the Introduction [32-36], the present work offers further insights into the behaviour and analysis of Brownian functionals (particularly those involving the Ornstein-Uhlenbeck process) which may help prompt other lines of enquiry.

To explain why the first passage area functional is asymptotically normal in the limit $\alpha \to \infty$, recall that $\mathcal{A}$ is the integral of a Gaussian process $x(\tau)$ up to time $\mathcal{T}$, which is random. It is this latter feature that makes finding the distribution of $\mathcal{A}$ challenging. As $\alpha \to \infty$, however, $\mathcal{T}$ is increasingly sharply defined around its mean value $\mathcal{T}_1(\alpha) = \log \alpha + O(1)$ (the variance of $\mathcal{T}$ is bounded as $\alpha \to \infty$; see (12, 13)). To leading order, therefore, one can assume all trajectories have the same duration in the limit. It is known that the integral $\mathcal{A}(t) = \int_0^t x(\tau)\,d\tau$ for a given $t$ has a Gaussian distribution with mean $\alpha[1 - e^{-t}]$ and variance $2t - 3 + 4e^{-t} - e^{-2t}$; see e.g. [42]. Making the association $t \approx \mathcal{T}_1(\alpha)$ the result follows, consistent with (18, 20).

The precise form of the area probability density remains an open question. By appealing to the governing differential equation which may be deduced from (3):

$$\left[\frac{\partial^2}{\partial \alpha^2} - \alpha \frac{\partial}{\partial \alpha}\right] P(\mathcal{A}|\alpha) = \alpha \frac{\partial}{\partial \mathcal{A}} P(\mathcal{A}|\alpha)$$

one can study the deviations from normality as $\alpha \to \infty$ as well as postulate the behaviour in the extreme tails as $\mathcal{A} \to 0$ (essential singularity, guided by the Brownian motion result [3, 16]; see also [39]) and $\mathcal{A} \to \infty$ (exponential decay, guided by the dominant singularity of the Laplace transform (4)):

$$-\log P(\mathcal{A}|\alpha) \sim \begin{cases} \dfrac{\alpha^3}{9\mathcal{A}}; & \mathcal{A} \to 0 \\[2ex] \phi \mathcal{A}; & \mathcal{A} \to \infty \end{cases}$$

where $\phi = 0.57279...$ is the smallest positive root of $D_{\phi^2}(-2\phi) = 0$ [21]. In the absence of a full solution, more formal confirmation and extension of these asymptotes would be a welcome next step. One approach might be to adapt the methods developed in [16], but we do not pursue this any further here.

**Appendix: Asymptotic behaviour of key integrals**

Following [39], to evaluate the limit $\alpha \to \infty$ of various integrals appearing in the main text, it is helpful to consider

$$I(\alpha, \beta) = \int_0^\infty e^{-\frac{1}{2}u^2} \left[ \frac{1-e^{-\alpha u}}{u} \right] u^\beta \, du; \qquad \beta > 0. \qquad (A1)$$

This may be written as

$$I(\alpha, \beta) = \int_{-\infty}^\infty e^{-u^2/2\alpha^2} \left[ \frac{1-e^{-u}}{u} \left( \frac{u}{\alpha} \right)^\beta \mathbf{1}_{u>0} \right] du$$

after which one can invoke the Plancherel formula to give

$$I(\alpha, \beta) = \frac{1}{\sqrt{2\pi}} \frac{\Gamma(\beta)}{\alpha^{\beta-1}} \int_{-\infty}^\infty e^{-\alpha^2 \omega^2/2} \left[ (i\omega)^{-\beta} - (1+i\omega)^{-\beta} \right] d\omega.$$

It is convenient to write this as the sum of two parts $I = I_1 + I_2$ defined by

$$I_1(\alpha, \beta) = \frac{1}{\sqrt{2\pi}} \frac{\Gamma(\beta)}{\alpha^{\beta-1}} \int_{-\infty}^\infty e^{-\alpha^2 \omega^2/2} \left[ (i\omega)^{-\beta} - 1 \right] d\omega$$

$$I_2(\alpha, \beta) = \frac{1}{\sqrt{2\pi}} \frac{\Gamma(\beta)}{\alpha^{\beta-1}} \int_{-\infty}^\infty e^{-\alpha^2 \omega^2/2} \left[ 1 - (1+i\omega)^{-\beta} \right] d\omega.$$

The integral $I_1$ can be evaluated exactly:

$$I_1(\alpha, \beta) = 2^{\frac{1}{2}\beta - 1}\Gamma\left(\frac{\beta}{2}\right) - \Gamma(\beta)\alpha^{-\beta}. \tag{A2}$$

Regarding $I_2$, as $\alpha \to \infty$ the dominant contribution comes from near $\omega = 0$, which permits one to expand around this point to derive

$$I_2(\alpha, \beta) = \frac{\Gamma(2+\beta)}{2\alpha^{2+\beta}}\left[1 + O\left(\frac{1}{\alpha^2}\right)\right]. \tag{A3}$$

By inspection, the integral $I_1$ determines the leading order behaviour of the full integral $I(\alpha, \beta)$ as $\alpha \to \infty$.

With reference to (A1), the Taylor expansion around $\beta = 0$ yields

$$I(\alpha, \beta) = \int_0^\infty e^{-\frac{1}{2}u^2}\left[\frac{1 - e^{-\alpha u}}{u}\right]du + \beta\int_0^\infty e^{-\frac{1}{2}u^2}\left[\frac{1 - e^{-\alpha u}}{u}\right]\log u\, du + \ldots$$

Based on the corresponding Taylor expansion of (A2) and (A3) one then has

$$\int_0^\infty e^{-\frac{1}{2}u^2}\left[\frac{1 - e^{-\alpha u}}{u}\right]du = \log\alpha + \frac{\gamma + \log 2}{2} + o(1) \tag{A4}$$

$$\int_0^\infty e^{-\frac{1}{2}u^2} \left[ \frac{1-e^{-\alpha u}}{u} \right] \log u \, du = -\frac{(\log \alpha)^2}{2} - \gamma \log \alpha$$

$$-\frac{1}{8}(\gamma + \log 2)(3\gamma - \log 2) - \frac{\pi^2}{16} + o(1).$$

(A5)

Further, after separating out (A1) into its constituent parts one has

$$I(\alpha, \beta) = 2^{\frac{1}{2}\beta - 1} \Gamma\left(\frac{\beta}{2}\right) - \int_0^\infty u^{\beta - 1} e^{-\frac{1}{2}u^2} e^{-\alpha u} \, du.$$

By comparison with (A2) and (A3) one then has the result

$$\int_0^\infty u^{\beta - 1} e^{-\frac{1}{2}u^2 - \alpha u} \, du = \frac{\Gamma(\beta)}{\alpha^\beta} \left[ 1 + O\left(\frac{1}{\alpha^2}\right) \right]; \qquad \alpha \to \infty. \qquad (A6)$$